
%
\input phyzzx.tex
\overfullrule=0pt
%
%

\def\RREF#1#2{\gdef#1{\REF#1{#2}#1}}
\def\jnl#1&#2(#3){\begingroup\let\jnl=\dummyj@urnal\sl #1\bf#2\rm
    (\afterassignment\j@ur\count255=#3)\endgroup}	
%
\def\PRL{ {\sl Phys. Rev. Lett.}   }

\def\NP { {\sl Nucl. Phys.}        }
\def\PL { {\sl Phys. Lett.}        }

\RREF\brez{E. Brezin and V. Kazakov, \PL {\bf B236} (1990) 144.}
\RREF\shen{M. Douglas and S. Shenker, \NP {\bf B335} (1990) 635.}
\RREF\gros{D. Gross and A.A. Migdal, \PRL {\bf 64} (1990) 127;
\NP{\bf B340} (1990) 333. }
\RREF\doug{M. Douglas, \PL {\bf B238} (1990) 176. }
\RREF\bank{T. Banks, M. Douglas, N. Seiberg and S. Shenker, \PL
{\bf B238} (1990) 279. }
\RREF\watt{S. Dalley, C.V. Johnson, T.R. Morris and A. Watterstam,
Mod. Phys. Lett. {\bf A7} (1992) 2753. }
\RREF\migd{D. Gross and A.A. Migdal, \PRL {\bf 64} (1990) 717.}
\RREF\isin{E. Br\'ezin, M. Douglas, V. Kazakov and S. Shenker, \PL
{\bf B237} (1990) 43.}
\RREF\mart{E. Martinec, G. Moore and N. Seiberg, \PL {\bf B263}
(1991) 190.}
\RREF\stau{G. Moore, N. Seiberg and M. Staudacher, \NP {\bf B362}
(1991) 665.}
\RREF\akaz{V.A. Kazakov, \PL {\bf B237} (1990) 212. }
\RREF\kost{I.K. Kostov, \PL {\bf B238} (1990) 181. }
\RREF\crmo {\u C. Crnkovi\'c, P. Ginsparg and G. Moore, \PL {\bf B237}
(1990) 196.}
\RREF\gelf{I.M. Gelfand and L.A. Dikii, Funct. Anal. Appl.
{\bf 11} (1977) 93.}
\RREF\hou{L. Houart, \PL {\bf B295} (1992) 37. }
\RREF\kaza{V.A. Kazakov, \PL {\bf A119} (1986) 140; D.V. Boulatov and
V.A. Kazakov, \PL {\bf B186} (1987) 379. }
\RREF\goul{P. Ginsparg, M. Goulian, M.R. Plesser and J. Zinn-Justin, \NP
{\bf B342} (1990) 539.}
\RREF\zuck{B. Lian and G. Zuckerman \PL {\bf B254} (1991) 417. }
\RREF\john{C.V. Johnson, Princeton preprint IASSNS-HEP-93/5 (hep-th
9301112) .}
\RREF\itz{C. Itzykson and J.-B. Zuber J. Math. Phys. {\bf 21} (1980)
411; D. Bessis, C. Itzykson and J.-B. Zuber,Adv. Appl. Math. 1 (1980)
109. }
\RREF\difr{P. Di Francesco and D. Kutasov,  proceedings of the Cargese
Workshop "Random Surfaces and Quantum Gravity", May 27 to June 2, 1990,
Plenum Press, New York 1991. }

\def\ULB{\address{Service de Physique Th\'eorique \break
Universit\'e Libre de Bruxelles, Boulevard du Triomphe \break
      CP 225, B-1050 Bruxelles, Belgium}}
\nopubblock
\titlepage
\line{\hfil\vbox{
\hbox{ULB--TH--02/93}
\hbox{March 1993}
}}

\title{Explicit Resolution of an Integrable $c(4,3)$ Open String Theory}

\author{Laurent Houart\footnote{\dag}{Aspirant
FNRS.}\footnote{\sharp}{e-mail: lhouart@ulb.ac.be}} \ULB
\abstract
{ We study the two-matrix model which represents the sum over closed and
open random surfaces coupled to an Ising model.
The boundary conditions are characterized by the fact that the Ising
spins sitting at the vertices of the boundaries are all in the same
state. We obtain the string equation and discuss the results.  }
\endpage

Since the discovery of the double scaling limit\refmark{\brez, \shen,
\gros}, important progress has been achieved in the understanding of
open-closed string theories in $(p,q)$ conformal minimal model
backgrounds.

We have now a complete coherent description of the $(2m-1,2)$
conformal minimal models coupled to 2D gravity. These theories are
described by the KdV hierarchy associated with the lie algebra
$sl(2,C)$. Starting with the usual one-hermitian-matrix
model\refmark{\brez, \shen, \gros}, which describes the sum over closed
surfaces, one finds that  the KdV flow equations organize the operator
structure of the $(2m-1,2)$ theories. Supplying ``initial conditions"
given by the string equation yields a complete description of closed
string theories in the $(2m-1,2)$ backgrounds.\hfill \break
The operator content of the theories is fully understood.
Indeed, the infinite number of operators found by BRST analysis in the
Liouville description\refmark{\zuck}is also present in the KdV
formulation. The remaining KdV flows are identified studying the
macroscopic loops\refmark{\bank, \mart, \stau}. They correspond to
boundary operators\refmark{\mart} (the boundary length and the infinite
number of operators associated with it). The KdV hierarchy associated
to $sl(2,C)$ is thus the good framework to describe both closed and
open string theories in $(2m-1,2)$ conformal minimal backgrounds.\hfill
\break  Moreover, a direct study of the one-hermitian-matrix model
supplemented with a logarithmic potential\refmark{\akaz}-- having the
effect of adding surfaces with boundaries of finite extent in the
partition function-- has been performed in the double scaling
limit\refmark{\kost}. This leads to the generalization of the previous
string equation to the case of a non-zero open string coupling.\hfill
\break Finally, it has been shown\refmark{\watt} that the $sl(2,C)$ mKdV
models are a different description of the same open-closed string
theories.

The description of the general (p,q) models\foot{ we consider
$p>q$} coupled to 2D gravity is much more involved. Closed string
theories in these backgrounds have been studied. They are realized in
terms of multi-matrix models and are characterized in the
double scaling limit by the generalized KdV hierarchies
\refmark{\doug, \goul, \difr}. For a given $q$ the $sl(q,C)$
KdV hierarchy organizes the operator structure of the theory. Once more,
the spectrum of the theory contains the operators found in the
Liouville description but some remaining flows no longer
admit an interpretation in terms of boundary operators\refmark{\mart}.
Furthermore, the $sl(q,C)$ KdV hierarchy does not allow a
description of the boundary length.\hfill \break
Related to this problem is the issue of finding a string
equation describing open string theories in $(p,q)$ backgrounds.
In this case, when surfaces with boundaries of finite extent are added
to the partition sum, one has to pay attention to the boundary problem.
There are indeed $q-1$ order parameters in the theory and the boundary
conditions associated with them have to be fixed. \hfill \break
Recently a string equation for general open string theories in
$(p,q)$ backgrounds has been derived by Johnson \refmark{\john} in the
integrable model framework. This equation is obtained by studying
the $sl(q,C)$ generalization of a mapping which transforms the $sl(2,C)$
$\tau$-function characterizing the closed case into the $sl(2,C)$
$\tau$-function of the open case. However by making use of the
beautiful underlying mathematical structure and bypassing the matrix
model route some physical informations are lost. It is very hard to know
what are the boundary conditions associated with the string equation.

This letter is concerned with the boundary condition problem. Starting
with the discrete formulation, we work out an explicit example: the
open string theory in the $(4,3)$ conformal background with fixed
boundary conditions. In the spirit of ref.[9], using the orthogonal
polynomial method, we study the two-matrix model which represents the
sum over closed and open random surfaces coupled to an Ising model
(IM). The boundary conditions are characterized by the fact that the
Ising spins sitting at the vertices of the boundaries are all in the
same state. We obtain the string equation and discuss the results.
\vskip1.5cm

We are going to study, in the double scaling limit, the following
two-matrix model: $F=ln Z$
$$Z=\int {\cal D}M_1 {\cal D}M_2 \exp{(-{1 \over \epsilon}
TrV(M_1,M_2))} \eqn\eqi$$
where $V$ is the asymmetrical potential:
$$V(M_1,M_2)={1 \over 2} M^2_1 -{1 \over 4} M^4_1 +
{1 \over 2} M^2_2 -{1 \over 4} M^4_2 - cM_1 M_2 + \gamma
\ln(1-e^{\mu}M^2_1) \eqn\eqii$$
and $\epsilon ={g \over N}$.\hfill \break
When $\gamma =0$, $F$ corresponds to the sum over closed surfaces
coupled to an Ising model\refmark{\kaza}. The two matrices $M_1,M_2$
are used to represent the two spin states. The solution of the model
in this case is well-known\refmark{\crmo, \isin, \migd}. Since we have
written $g^{-1}$-- the parameter related to the bulk cosmological
constant -- in front of the potential, it is the direct diagramatic
expansion of \eqi \  which represents the sum over random surfaces
($\epsilon$ is the closed string coupling) and the Ising spins sit on
the vertices of the triangulations. When $\gamma \not= 0$ the
logarithmic potential has the effect of adding surfaces with arbitrary
boundaries of finite extent to the partition
function\refmark{\akaz,\kost}. In this case, $F$ represents the sum
over closed and open triangulated surfaces with the Ising spins sitting
at the vertices of the boundaries all in the same state (e.g. all
up). The parameter $\gamma$ is the open string coupling. In the
continuum limit $\mu$ will correspond to the parameter which couples
to the boundary length as well as the parameter which couples to the
boundary magnetization (there is no way to distinguish them in this
model).

The model can be solved in the double scaling limit using the
well-known orthogonal polynomial method\refmark{\itz}.
We introduce as usual the orthogonal polynomials defined by:
$P^{\pm}_n(\alpha) = \alpha^n + O(\alpha^{n-1})$
$$\int d\Theta(\lambda,\nu) P^{-}_n(\lambda) P^{+}_m(\nu)=
h_n \delta_{m,n} \eqn\eqiii$$
where $d\Theta(\lambda,\nu)=d\lambda  d\nu \exp{(-{1 \over \epsilon}
V(\lambda,\nu))}$. \hfill \break
We have $F= {\sum}_n (N-n) log f_n$ where $f_n={h_n
\over h_{n-1}}$.\hfill \break
Only $P^{-}_n(\lambda)$ satisfies a three term recursion relation. We
have:
$$\lambda P^{-}_n(\lambda)=P^{-}_{n+1}(\lambda)+r^-_n
P^{-}_{n-1}(\lambda) + s^-_n P^{-}_{n-3}(\lambda)\eqn\eqxxxx $$
$$\nu P^{+}_n(\nu)=P^{+}_{n+1}(\nu)+r^+_n P^{+}_{n-1}(\nu) + s^+_n
P^{+}_{n-3}(\nu)+{\sum}^{n-4}_0 \alpha^{+}_k P^{+}_k(\nu) \eqn\eqiv$$
It's convenient to introduce the orthonormal basis:\hfill \break
$\mid n \rangle ={P^{-}_n(\lambda) \over \sqrt{h_n}} \quad , \quad
\langle n \mid ={P^{+}_n(\nu) \over \sqrt{h_n}}$\hfill \break
and to define:\hfill \break
$t^+_n = {r^+_n+r^-_n \over 2} \quad ,\quad t^-_n = {r^+_n-r^-_n \over
2}$ \hfill \break
The functions $t,s,f$ satisfy:
$$c t^+_n = f_n(1- t^+_{n+1}- t^+_n- t^+_{n-1})- \gamma
f^{1 \over 2}_n K(n,n-1) \eqn\eqv$$
$$c t^-_n = f_n(t^-_{n+1}+ t^-_n + t^-_{n-1})- \gamma
f^{1 \over 2}_n K(n,n-1) \eqn\eqvi$$
$$c s^+_n = -f_nf_{n-1}f_{n-2}-2 \gamma
\sqrt{f_nf_{n-1}f_{n-2}} K(n,n-3) \eqn\eqvii$$
$$c s^-_n = -f_nf_{n-1}f_{n-2} \equiv c s_n \eqn\eqvii$$
where
$$K(n,m) \equiv \int d\Theta(\lambda,\nu) {P^{+}_n(\nu) \over
\sqrt{h_n}}  ({\lambda \over e^{-\mu} - \lambda^2})
{P^{-}_m(\lambda) \over \sqrt{h_m}}= \langle n \mid
{\lambda \over e^{-\mu} - \lambda^2}
\mid m \rangle \eqn\eqviii$$
The string equation is given by:
$$\eqalign{ \epsilon n  & = -cf_n
+t^+_n-t^+_n(t^+_{n+1}+ t^+_n+ t^+_{n-1})- t^-_n(t^-_{n+1}+ t^-_n+
t^-_{n-1}) \cr & - s_n- s_{n+1}-s_{n+2}- \gamma  f^{1 \over 2}_n
K(n-1,n)+{\gamma \over c} A(n) \cr} \eqn\eqix$$
$$\eqalign{ 0 & = t^-_n-t^+_n(t^-_{n+1}+ t^-_n+ t^-_{n-1})-
t^-_n(t^+_{n+1}+ t^+_n+t^+_{n-1}) \cr & + \gamma f^{1 \over 2}_n
K(n-1,n)+{\gamma \over c} A(n) \cr} \eqn\eqx$$
where
$$\eqalign {A(n) &= \sqrt{f_nf_{n-1}f_{n-2}} K(n,n-3)+
\sqrt{f_{n-1}f_nf_{n+1}} K(n+1,n-2) \cr &+ \sqrt{f_nf_{n+1}f_{n+2}}
K(n+2,n-1) \cr } \eqn\eqxi$$

We now analyse separately the cases $\gamma = 0$ and $\gamma \not= 0$.

$\bullet \qquad \gamma = 0$

It corresponds to the closed case which has already been
solved\refmark{\crmo, \migd, \isin}. We rederive briefly the known
results using our notations. The continuum limit is characterized
by: \hfill \break $f_c={1 \over 12},\quad t_c={1 \over 6},\quad c={1
\over 4}, \quad g_c={10 \over (12)^2}$ \hfill \break
As usual, we define the scaling variable $z$
by $n\epsilon=g_c(1-a^2z)$ and we take the scaling functions:
$$\eqalign{ f & =f_c(1-a^{{2 \over 3}}u) \cr
t^+ &= t_c(1+a^{{2 \over 3}}r_1+a^{{4 \over 3}}r_2+a^2 r_3) \cr
t^- & =t_caw+0(a^{{5 \over 3}}) \cr} \eqn\eqxxxxx$$
The double scaling limit is reached when $N \rightarrow \infty$ and
$a \rightarrow 0$ with ${1 \over Na^2}=G a^{{1 \over 3}}$ held fixed.
The parameter $G$ is the renormalized closed string coupling. The
susceptibility is given in this limit by  ${d^2 \over dx^2}F=-{1 \over
G^2}u(x)$ where $x=a^{-2}(1-g/g_c)$ is the continuum bulk cosmological
constant.\hfill \break
The first non-zero contribution of eq.\eqvi
\ is of order $a^{{5 \over 3}}$ and gives:    $${\cal R}_0 \equiv G^2
w^{\prime \prime}-3uw = 0 \eqn\eqxii$$  where prime stands for ${d
\over dz}$. The string equation \eqix \  becomes (at order $a^2$):
$${\cal R}_1 \equiv {6 \over 5} \lbrack {1 \over 3} u^3+{8 \over
(12)^2}G^4 u^{(4)}-{1 \over 2}G^2 u^{\prime \prime}u- {1 \over 4}G^2
(u^\prime)^2+w^2 \rbrack-z = 0\eqn\eqxiii$$  One verifies that the
equation \eqx \ is satisfied.  The operator $\lambda$ becomes:
$\lambda=\lambda_c \lbrack  1+aQ+0(a^{{4\over 3}}) \rbrack$ with
$\lambda_c={8 \over 3 \sqrt{12}}$ and
$$Q=-{1 \over 2}G^3 \partial^3_z +{3 \over 8}G \lbrace u,\partial
\rbrace - {3 \over 4}w \eqn\eqxiv$$

$\bullet \qquad \gamma \not= 0$

Considering $K(n,m)$, in the double scaling limit we have:
$$K(n,n+i)={\lambda_c \over Na^2} \langle z \mid
{1 \over e^{-\mu}- \lambda^2} \mid z-{i \over Na^2} \rangle \quad
\lbrack 1 +O(a^{{2 \over 3}}) \rbrack \eqn\eqxv$$
We make the hypothesis that $Q$ derived from eq.\eqxxxx \ and given by
eq.\eqxiv \ is not modified at order $a$ when $\gamma \not= 0$. It will
be checked later that this is indeed the case.\hfill \break
Choosing $\exp{(-\mu_c)}=\lambda^2_c$ and defining the renormalized
boundary ``cosmological" constant $M$ by:
$$\mu_c-\mu = aM \eqn\eqxvi$$
eq.\eqxv \  becomes:
$$\eqalign{ K(n,n+i) & ={G a^{-{2 \over 3}} \over
\lambda_c }  \langle z \mid {1 \over M-2Q} \mid z-{i \over Na^2}
\rangle \quad \lbrack 1 +O(a^{{2 \over 3}}) \rbrack \cr
& = {G a^{-{2 \over 3}} \over \lambda_c} \lbrack R_0(z,M) -ia^{{1 \over
3}}G R_1(z,M) +O(a^{{2\over 3}}) \rbrack  \cr} \eqn\eqxvii$$
where $R_0(z,M)$ is the resolvent and $R_1(z,M)$ is the first jet of the
resolvent\refmark{\gelf}, to wit:

$$\eqalign{ & R_0(z,M) \equiv \langle z \mid {1 \over M-2Q} \mid
z \rangle \cr &  R_1(z,M) \equiv \partial_{\tilde z} \langle z \mid {1
\over M-2Q} \mid \tilde z \rangle \biggl \vert_{\tilde z= z} \cr
}\eqn\eqxviii$$
We now turn to the determination of the string equation. Using the
definition \eqxxxxx \ of $w$, eq.\eqvi \  leads to:
$$0= a^{{5 \over 3}}f_c t_c(G^2w^{\prime \prime}-3uw)- \gamma
f^{{1 \over 2}}_c \lbrack {G a^{-{2 \over 3}} \over \lambda_c} R_0(z,M)
+ O(a^{-{1 \over 3}}) \rbrack \eqn\eqxxx $$
It is thus natural to define the renormalized open string coupling
$\Gamma$ by
$$\gamma = a^{{7 \over 3}} \Gamma \eqn\eqxix$$
As expected\refmark{\akaz, \kost} $\gamma$ scales like ${1 \over
N}$.
Eq.\eqxxx \ becomes:
$${\cal R}_0=27  \Gamma G R_0(z,M) \eqn\eqxx$$
Let us consider now eq.\eqv. Writing $t^+=t^+(\gamma =0)+ {\tilde
t}^+$ one finds that :
$${\tilde t}^+=-2f^{{1\over 2}}_c
\lambda^{-1}_c \Gamma \lbrack  a^{{5 \over 3}} R_0+ a^2 R_1+ O(a^{{7
\over 3}}) \rbrack \eqn\eqxxi $$
Replacing $t^+$, $t^-$ and $s$ in the string
equation \eqix \ we see that the terms of order $a^{{5 \over
3}}$ cancel and we obtain:
$$\eqalign{{\cal R}_1 &= {G^2 \Gamma f^{1 \over
2}_c \over g_c \lambda_c} \lbrack ({9f_c \over c}+1)R_1(z,M)- (1+{3f_c
\over c})R^\prime_0(z,M)  \rbrack \cr & = {54 \over 5}G^2 \Gamma \lbrack
2R_1(z,M)-R^\prime_0(z,M) \rbrack \cr }\eqn\eqxxii $$
Moreover eq.\eqx \ is satisified and the hypothesis that $Q$
is still given by eq.\eqxiv \ is right, the corrections being indeed of
order $O(a^{{5 \over 3}})$.

Now, using a technique developed by Gel'fand and Dikii\refmark{\gelf},
we compute a system of equations for the resolvent and the first five
jets (see theorem 6 of ref.[20]). We then extract a system of two
equations for $R_0$ and $R_1$. The result is:
$$\eqalign{ &
-G^3(2R_1-R^\prime_0)^{(3)}+{3 \over 4}G u^\prime
(2R_1-R^\prime_0) + {3 \over 2}G u(2R_1-R^\prime_0)^\prime \cr
& = 3 w^\prime R_0 +{9 \over 2}wR^\prime_0 +3 M R^\prime_0 \cr
}\eqn\eqxxiii$$
$$\eqalign{ & {1 \over 9}G^4 R^{(5)}_0 -{5 \over 6} G^2
 u R^{(3)}_0- {5 \over 4}G^2u^\prime R^{(2)}_0-({3 \over 4}
G^2 u^{\prime \prime}  - u^2) R^\prime_0   - ({1 \over 6}
G^2 u^{(3)}  -u u^\prime) R_0 \cr  & = {1 \over 2}G
w^\prime (2R_1-R^\prime_0) +{3 \over 2} G w
(2R_1-R^\prime_0)^\prime+G M (2R_1-R^\prime_0)^\prime \cr }
\eqn\eqxxiv$$

The model \eqi \ is thus completely solved in the double scaling
limit. The solution is given by eqns.\eqxx, \eqxxii, \eqxxiii \ and
\eqxxiv.

The results we obtain are compatible with those of ref.[14].
Indeed, modulo rescalings we recover in the particular (4,3) case the
string equation found by Johnson (in his notation: ${\hat
R}^2=2R_1-R^\prime_0$ and ${\hat R}^3=2R_0$). The
results we obtain confirm also that the $sl(3,C) \quad mKdV$ models are
an equivalent description of open string theories in (p,3)
backgrounds\refmark{\hou}.

Finally, it would be interesting to solve in the same spirit the model
with the logarithmic potential:
$\ln(1-e^{\mu+\rho}M^2_1)+\ln(1-e^{\mu-\rho}M^2_2)$.
This model would allow, in the double scaling limit, a description of
the boundary magnetization constant $\rho$ as well as a description of
the boundary cosmological constant $\mu$. It would presumably bring new
insight to the boundary cosmological constant problem\refmark{\mart}.

\refout
\endpage
\bye